\documentclass[11pt, oneside,reqno]{amsart}

\usepackage[top=2.5cm, bottom=2.5cm, left=2.5cm, right=2.5cm]{geometry}

\usepackage[english]{babel}

\usepackage[T1]{fontenc}

\usepackage{textcomp}

\usepackage[utf8]{inputenc}

\usepackage{amsfonts, amsmath, amssymb, mathtools, amsthm}
\usepackage[makeroom]{cancel}
\usepackage[none]{hyphenat}
\usepackage{bigints}
\usepackage{xcolor}

\usepackage{siunitx}
\usepackage{fancyhdr}
\usepackage{graphicx}
\usepackage{verbatim}
\usepackage{xcolor}
\usepackage{pgf, tikz, tkz-base}
\usepackage{mathrsfs}
\usepackage{tkz-euclide}
\usepackage{xfp}

\usetikzlibrary{shapes, calc, shapes, arrows, babel}
\usepackage[margin=16pt,font=small,labelfont=bf]{caption} 

\usepackage[maxbibnames=99, sorting=nyt]{biblatex}
\usepackage{parskip}

\usepackage[colorlinks=true]{hyperref}

\begin{document}

\title
{Tides and dumbbell dynamics}

\author{
Benedetto Scoppola$^{1}$ \and
Matteo Veglianti$^{2}$\and 
Alessio Troiani$^{3}$}


\maketitle

\begin{center}
{\footnotesize
\vspace{0.3cm}$^{1}$  Dipartimento di Matematica,\\
Universit\`a di Roma
``Tor Vergata''\\
Via della Ricerca Scientifica - 00133 Roma, Italy\\
\texttt{scoppola@mat.uniroma2.it}\\

\vspace{0.3cm} $^{2}$ Dipartimento di  Fisica,\\ Universit\`a di Roma
``Tor Vergata''\\
Via della Ricerca Scientifica - 00133 Roma, Italy\\
\texttt{matteoveglianti@roma2.infn.it}\\ 

\vspace{0.3cm}$^{3}$  Dipartimento di Matematica “Tullio Levi–Civita”,\\ Università degli Studi di Padova,
\\Via Trieste, 63, 35131 Padova, Italy\\
\texttt{alessio.troiani@unipd.it}\\
}

\end{center}

\begin{abstract}
We discuss a model describing the effects of tidal dissipation on the satellite's orbit
in the two body problem. Tidal bulges are described in terms of a dumbbell, coupled to the rotation by a dissipative interaction. The assumptions on this dissipative coupling turns out to be crucial in the evolution of the system.
\end{abstract}

\section{INTRODUCTION}
Celestial bodies such as the Earth or some of the Jovian satellites consists of several layers
of materials with different density and interacting differently with other celestial body
(e.g. the Moon, the Sun, Jupiter or the other satellites).
As a consequence of this interaction and the non homogeneity of the inner structure,
tidal bulges are created on the surface of these celestial bodies. 
The rotational velocity of the bulges is, in general, not the same of that of the inner layers.
It is reasonable to think that the friction between the bulges and the 
remainder of the body depends on their mutual angular velocity.

In this framework we introduce a simple model in order to compute the effects of tidal dissipation in the two body problem. In particular we want to investigate the effects that this dissipation has on the orbits of both celestial bodies.
We will assume that the two bodies have very different masses, say $M\gg m$, and we call {\it planet} the 
body with mass $M$ and {\it satellite} the body with mass $m$. 
To keep the model as simple as possible
we assume the axes of rotation of both the planet and the satellite to be perpendicular to the orbital 
plane. We study the effects of the tides formed  by the satellite on the planet and, with the same 
approach, the effects of the tides formed by the planet on the satellite, assuming that the satellite 
is in $1:1$ spin-orbit resonance with the planet.

In our model, we describe the system in terms of  the dumbbell dynamics: the planet and the satellite are described in terms of a point $P$ of mass $M-\mu$ and a mechanical dumbbell centered in $P$, i.e., a system of two points, each having mass $\mu/2$, constrained to be at fixed mutual distance $2r$, having $P$ as center of mass. The idea is to substitute the study of the two tidal bulges with the study of their respective centers of mass.  
There is a vast literature concerning the study of the shape and the rheological properties of the celestial body.
In these works, following Darwin's classical approach, the tidal potential is
written in terms of its Fourier components each having its own rheological parameters,
e.g. its own dissipation function. Relevant references include
\cite{efroimsky2012bodily, efroimsky2013tidal, efroimsky2015tidal, ferraz2015dissipative, gevorgyan2020andrade}.
Though a detailed analysis of this kind allows to study the properties of
celestial bodies in terms of theory of viscoelastic bodies, our simple
model has a different focus:
we set aside a detailed discussion concerning the viscoelastic 
inner structure of the bodies,
but, nevertheless, we take into account the effects of the tidal torque 
exerted on the bulges, obtaining for instance in a clear way, its classical 
expression, see appendix below. 
This approach helps to clarify that the classical expression of the
tidal torque may be derived independently of the model of friction chosen
to describe the rheology of the celestial body.
We will assume that the motion of the tidal dumbbell and the rotation of the related heavenly body (planet or satellite) are coupled by a viscous dissipative friction, and we will use the simplest friction model involving an explicit dependence on the angular velocities. 

With this assumptions we will 
use an approach based on the Rayleigh dissipation function, 
obtaining in a unified and relatively standard way the equations of motions in the two cases mentioned above. We will call the study of the influence on the satellite's orbit of the tides on the planet 
{\it Earth--Moon system}, while the study of the effect of the tides on the satellite, supposed in $1:1$ resonance, will be called {\it Jupiter--Io system}. All the computations will be performed to the lowest order in the small parameters (the ratio between the radius of the bodies and the orbit of the satellite and the eccentricity).

The effects of tidal dissipation on the orbital parameters of both bodies
are, in this model, quite clear.
We hope that the model will serve as a stepping stone for the study of
the effects of tidal dissipation in systems involving several celestial objects
such as the case of Jovian satellites where resonance and tidal dissipation
seem to be closely related.

Note that the dynamics of the dumbbells in celestial mechanics has been already studied in different contexts as in \cite{CellettiSidorenko} and \cite{sidorenko2010spring} in which the dumbbell is
used to study satellites' attitudes. To our knowledge, however, the idea of applying 
the dumbbell dynamics to the dissipative tidal effects is not present in literature yet.

The work is organized as follows. In section~2 we present the equations of motions of the Earth--Moon system, i.e., considering the tidal torque exerted by the satellite on a dumbbell centered on the planet. Then we evaluate the evolution of the orbital parameter of the Moon due to this interaction 
and show that the dissipation tends to circularize the orbit of the Moon. In section~3 we study the Jupiter--Io system, showing that also in this case the orbit tend to be circularized. In both cases we 
prove that the eccentricity tends to zero exponentially. Finally section~4 is devoted to future developments of this approach.
In the appendix we compute directly the torque between the dumbbell and the other body.

\section{EARTH--MOON SYSTEM.}

\begin{figure}[!ht]
    \includegraphics[width=0.7\textwidth]{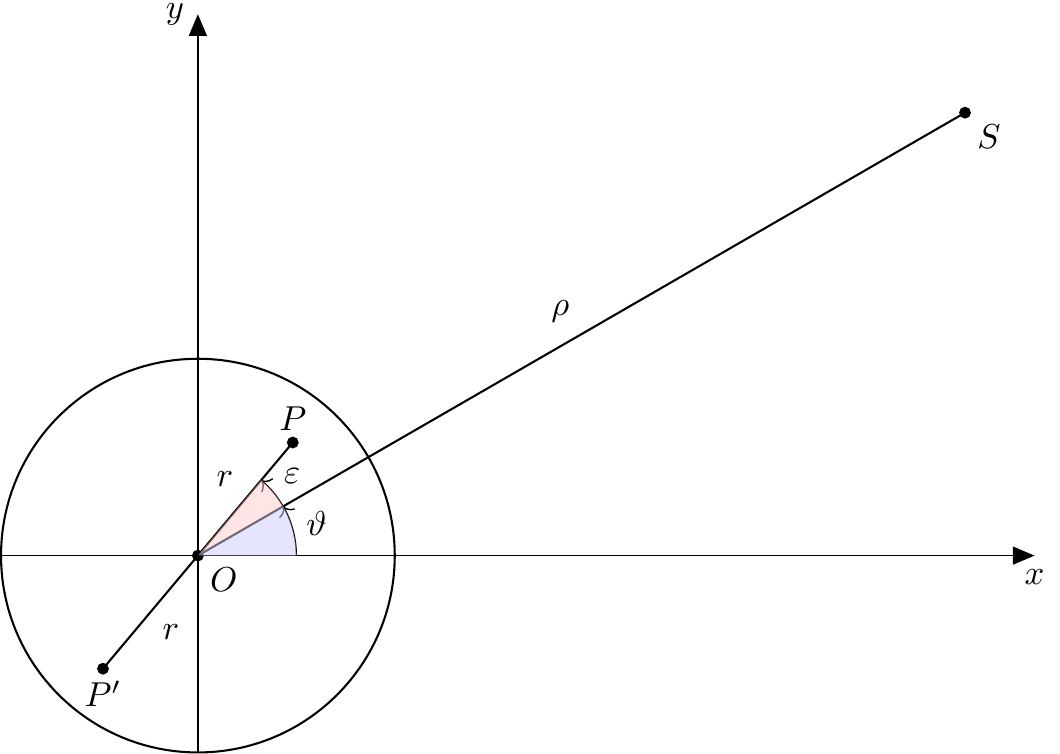}
    \caption{Earth--Moon system}
    \label{fig:earth_moon_system}
\end{figure}

In this section we want to study the evolution of the Earth--Moon system. We derive the equations of motion in a Lagrangian formalism.\\
To this end, as we show in Fig.~\ref{fig:earth_moon_system}, we imagine the Earth, of total mass $M_E$, as a sphere of radius $R_E$ plus a symmetric dumbbell of diameter $2r$ and mass $\mu$.
$\mu$ depends on the semi-major axis $a$ of the orbit of the satellite that varies in time, as we show in appendix, see \eqref{mu} and \eqref{rho_time}. However, in our model we consider $\mu$ constant. This assumption is justified a posteriori by the fact that we are interested to study the effects of dissipation on orbital parameters, such as, for example, the circularization of the orbit. 
These effects have characteristic timescales that are faster than the characteristic timescale of elongation of the semi-major axis. Hence, if we study the evolution of the system for a time 
much shorter than the latter timescale,
we can consider $a$, and $\mu$ as constants. 
The calculation concerning the comparison between the characteristic time of the circularization of the orbit and the characteristic time of elongation of the semi-major axis is presented in the appendix.
We imagine the Moon, indicated with $S$, as a point of mass $m$. We indicate with $\rho$ the distance between the Moon and the center of the Earth: we observe that in an elliptical orbit this distance varies with time. Finally, we call $\varphi$, $\vartheta$ and $(\vartheta + \varepsilon)$ the angular positions of the Earth, the Moon and the dumbbell with respect to a fixed direction ($x$-axis), respectively. So $\dot{\varphi}$, $\dot{\vartheta}$ and $(\dot{\vartheta} + \dot{\varepsilon})$ are the angular velocities of the Earth, the Moon and the dumbbell respectively.\\ Being $M_E\gg m$ we will assume the Earth to be fixed at the origin of the reference system.
The total kinetic energy is the sum of the kinetic energies of the Moon, the Earth and the dumbbell:
\begin{equation}
    \label{kinetic}
    \mathcal{T} = \frac{1}{2} m (\dot{\rho}^2 + {\rho}^2 \dot{\vartheta}^2) + \frac{1}{2} I \dot{\varphi}^2 + \frac{1}{2} \mu r^2 (\dot{\vartheta} + \dot{\varepsilon})^2
\end{equation}
where $I$ is the Earth's moment of inertia.

The potential energy is the sum of three pieces of gravitational attraction: that between the Earth (deprived of the dumbbell) and the Moon; that between point $P$ (a bulge of the dumbbell) and the Moon and that between point $P'$ (the other bulge of dumbbell) and the Moon:
\begin{equation}
    \label{potential}
    \mathcal{V} = -\frac{k(M_E-\mu)m}{\rho} - \frac{k\frac{\mu}{2}m}{\sqrt{r^2+{\rho}^2-2r \rho \cos{\varepsilon}}} - \frac{k\frac{\mu}{2}m}{\sqrt{r^2+{\rho}^2+2r \rho \cos{\varepsilon}}}. 
\end{equation} 
where $k$ is the universal gravitational constant.
If we now expand the potential up to the second order in $\frac{r}{\rho}$ (that is a dimensionless small parameter), we obtain the following expression:   
\begin{equation}
    \label{potential_second}
    \mathcal{V} = -\frac{gm}{\rho} \left[1 + \frac{\mu}{M_E} \frac{r^2}{{\rho}^2} \left( \frac{3}{2} \cos^2{\varepsilon} - \frac{1}{2}\right) \right], 
\end{equation} 
with $g=kM_E$.\\
So, the Lagrangian of the system is, with all the aforementioned assumptions:
\begin{equation}
    \label{Lagrangian}
    \mathcal{L} = \mathcal{T} - \mathcal{V} = \frac{1}{2} m (\dot{\rho}^2 + {\rho}^2 \dot{\vartheta}^2) + \frac{1}{2} I \dot{\varphi}^2 + \frac{1}{2} \mu r^2 (\dot{\vartheta} + \dot{\varepsilon})^2 + \frac{gm}{\rho} \left[1 + \frac{\mu}{M_E} \frac{r^2}{{\rho}^2} \left( \frac{3}{2} \cos^2{\varepsilon} - \frac{1}{2}\right) \right].
\end{equation}\\
Now we want to take into account the dissipation of energy. The reasonable mechanism of such dissipation arise from the fact that a friction between the dumbbell and the underlying Earth is present. Since both the ocean and the Earth can be considered fluids (by assuming the Earth a highly viscous fluid), is reasonable to assume a Stokes-type friction both for the oceanic and the solid tides: namely a friction proportional to the difference between the angular velocity of the Earth (that is $\dot{\varphi}$) and that of the ocean's bulges (that is $(\dot{\vartheta} + \dot{\varepsilon})$).\\  
So we assume a frictional force \footnote{More precisely, a frictional torque} of the form:
\begin{equation}
    \label{frictional_force}
    f=-\alpha (\dot{\varphi} - \dot{\vartheta} - \dot{\varepsilon}),
\end{equation}
with $\alpha$ a small friction coefficient. Although $\alpha$ depends on $\mu$, we can consider it constant as long as we consider $\mu$ constant too, as we argued above.\\
A standard approach to treat a viscous friction in Lagrangian formalism is to use the Rayleigh's dissipation function $R$, defined as the function such that $\frac{\partial R}{\partial \dot{q}_i} = f_i$, where $f_i$ is the frictional force acting on the $i$-th variable.\\
In our case, the Rayleigh's dissipation function assumes the form:
\begin{equation}
    \label{Rayleigh}
    R=-\frac{1}{2}\alpha \left( \dot{\varphi} - \dot{\vartheta} - \dot{\varepsilon} \right)^2.
\end{equation}
The Euler-Lagrange equations become:
\begin{equation}
    \label{E_L_eq}
    \frac{d}{dt} \left( \frac{\partial \mathcal{L}}{\partial \dot{q}_i} \right) = \frac{\partial \mathcal{L}}{\partial q_i} + \frac{\partial R}{\partial \dot{q}_i}.
\end{equation}
Finally, it is easy show that, being $R$ of the form $R=-\frac{1}{2} v^2(\dot{q})$ with $v(\dot{q}) = \sum_j a_j \dot{q}_j$ (that is $v(\dot{q})$ linear in $\dot{q}$),  the energy dissipation rate is:
\begin{equation}
    \label{en_diss_rate}
    \dot{E} = 2R = -\alpha \left( \dot{\varphi} - \dot{\vartheta} - \dot{\varepsilon} \right)^2.
\end{equation}
Notice that in the classical literature, see for instance \cite{MurrayDermott}, (4.151), the dissipation is assumed to be linear in the difference of angular  velocities, implying a non differentiable behavior in $\dot{\varphi} - \dot{\vartheta} - \dot{\varepsilon}$.

The Euler-Lagrange equations \eqref{E_L_eq} leads to the following equations for the dynamical variables $\rho$, $\vartheta$, $\varphi$, $\varepsilon$:
\begin{equation}
    \label{rho}
    m \Ddot{\rho} = \frac{\partial \mathcal{L}}{\partial \rho},
\end{equation}
\begin{equation}
    \label{theta}
    \frac{d}{dt} \left( m \rho^2 \dot{\vartheta} \right) + \mu r^2 \left( \Ddot{\vartheta} + \Ddot{\varepsilon} \right) = \frac{\partial R}{\partial \dot{\vartheta}} = \alpha \left( \dot{\varphi} - \dot{\vartheta} - \dot{\varepsilon} \right),
\end{equation}
\begin{equation}
    \label{phi}
    I \Ddot{\varphi} = \frac{\partial R}{\partial \dot{\varphi}} = - \alpha \left( \dot{\varphi} - \dot{\vartheta} - \dot{\varepsilon} \right),
\end{equation}
\begin{equation}
    \label{epsilon}
    \mu r^2 \left( \Ddot{\vartheta} + \Ddot{\varepsilon} \right) = \frac{\partial \mathcal{L}}{\partial \varepsilon} + \frac{\partial R}{\partial \dot{\varepsilon}} = \frac{\partial \mathcal{L}}{\partial \varepsilon} + \alpha \left( \dot{\varphi} - \dot{\vartheta} - \dot{\varepsilon} \right).
\end{equation}\\
From these four equations, we can write down two interesting relations.\\
First, from \eqref{theta} and \eqref{phi} we obtain the conservation of angular momentum:
\begin{equation*}
    \frac{d}{dt} \left( m \rho^2 \dot{\vartheta} \right) + \mu r^2 \left( \Ddot{\vartheta} + \Ddot{\varepsilon} \right) + I \Ddot{\varphi} = 0 \implies m \rho^2 \dot{\vartheta} + \mu r^2 \left( \dot{\vartheta} + \dot{\varepsilon} \right) + I \dot{\varphi} = J = const.
\end{equation*}
Second, from \eqref{theta} and \eqref{epsilon} we obtain the equation that determines the evolution of orbital angular momentum $J^{(O)}$:
\begin{equation}
    \label{L_orb_dot}
    \frac{d}{dt} \left( m \rho^2 \dot{\vartheta} \right) = - \frac{\partial \mathcal{L}}{\partial \varepsilon}.
\end{equation}
The explicit expression of $\frac{\partial \mathcal{L}}{\partial \varepsilon}$ and its comparison with the classical form of the tidal torque are briefly presented in the first part of the appendix.
\\

In equations \eqref{rho} to \eqref{epsilon}, there are some negligible terms. In fact, assuming small eccentricity, $\rho \sim a$, we observe that the results presented in the appendix, namely \eqref{Gamma}, imply that $\frac{\partial \mathcal{L}}{\partial \varepsilon} \propto \frac{\mu}{M_E} \frac{r^2}{a^3}$, and equation \eqref{L_orb_dot} becomes:
\begin{equation*}
    m a^2 \Ddot{\vartheta} \propto \frac{1}{a} \left( \frac{\mu}{M_E} \frac{r^2}{a^2} \right),
\end{equation*}
We want to study the system keeping the lowest order in $\frac{\mu}{M_E} \frac{r^2}{a^2}$, which is a very small quantity. Since:
\begin{equation*}
    \mu r^2 \Ddot{\vartheta} \propto \frac{M_E}{m} \frac{1}{a} \left( \frac{\mu}{M_E} \frac{r^2}{a^2} \right)^2
\end{equation*}
the term $\mu r^2 \Ddot{\vartheta}$ in \eqref{theta} and \eqref{epsilon} can be neglected. Moreover, it is also reasonable assume initial conditions such that $\Ddot{\varepsilon} = O ( \Ddot{\vartheta} )$, namely the variation of angular velocity of the bulges is of the same order than the variation of angular velocity of the Moon. Hence the term $\mu r^2 \Ddot{\varepsilon}$ in \eqref{theta} and \eqref{epsilon} can be neglected too.\\

Consequently, the simplified equations of motion are:
\begin{equation}
    \label{rho2}
    m \Ddot{\rho} = \frac{\partial \mathcal{L}}{\partial \rho},
\end{equation}
\begin{equation}
    \label{theta2}
    \frac{d}{dt} \left( m \rho^2 \dot{\vartheta} \right) = \alpha \left( \dot{\varphi} - \dot{\vartheta} - \dot{\varepsilon} \right),
\end{equation}
\begin{equation}
    \label{phi2}
    I \Ddot{\varphi} = - \alpha \left( \dot{\varphi} - \dot{\vartheta} - \dot{\varepsilon} \right),
\end{equation}
\begin{equation}
    \label{epsilon2}
    \frac{\partial \mathcal{L}}{\partial \varepsilon} + \alpha \left( \dot{\varphi} - \dot{\vartheta} - \dot{\varepsilon} \right) = 0.
\end{equation}\\
We call $G$ the orbital angular momentum of the Moon (that is a using Delaunay canonical variable):
\begin{equation}
    \label{Gdef}
    G=m \rho^2 \dot{\vartheta},
\end{equation}
from \eqref{theta2} we have:
\begin{equation}
    \label{G_dot}
    \dot{G}=\alpha \left( \dot{\varphi} - \dot{\vartheta} - \dot{\varepsilon} \right).
\end{equation}
Moreover, the rate of dissipation of energy is given by \eqref{en_diss_rate}. The energy variation of the system is made of three contributions: the variation of energy of the Earth, the variation of energy of the Moon (namely the variation of orbital energy $E^{(O)}$) and the variation of energy of the bulges. The latter can be neglected for the same reason why we have neglected the variation of the angular momentum of the bulges. So \eqref{en_diss_rate} becomes:
\begin{equation}
    \label{E_dot}
    -\alpha \left( \dot{\varphi} - \dot{\vartheta} - \dot{\varepsilon} \right)^2 = \frac{d}{dt} \left( \frac{1}{2} I \dot{\varphi}^2 \right) + \frac{d E^{(O)}}{dt},
\end{equation}
where $E^{(O)}$ is the orbital energy of the Moon:
\begin{equation*}
    E^{(O)} = \frac{1}{2} m (\dot{\rho}^2 + {\rho}^2 \dot{\vartheta}^2) - \frac{gm}{\rho}. 
\end{equation*}
From \eqref{phi2}, \eqref{G_dot} and \eqref{E_dot}, we have:
\begin{equation}
    \label{EO_dot}
    -\alpha \left( \dot{\varphi} - \dot{\vartheta} - \dot{\varepsilon} \right)^2 = -\alpha \dot{\varphi} \left( \dot{\varphi} - \dot{\vartheta} - \dot{\varepsilon} \right) + \frac{d E^{(O)}}{dt} \implies \frac{d E^{(O)}}{dt} = \left( \dot{\vartheta} + \dot{\varepsilon} \right) \dot{G}.
\end{equation}\\
Consider now the other canonical Delaunay variable $L$, defined as:
\begin{equation*}
    L = \frac{m^{\frac{3}{2}}g}{\sqrt{-2E^{(O)}}};
\end{equation*}
hence
\begin{equation}
    \label{L_dot}
    \dot{L} = \frac{g {{\dot E}^{(O)}}}{ \left( -\frac{2 E^{(O)}}{m}  \right)^{\frac{3}{2}} }.
\end{equation}
Now, on a Keplerian orbit, we have:
\begin{equation*}
    E^{(O)}=-\frac{mg}{2a} \implies -\frac{2E^{(O)}}{m} = \frac{g}{a}, 
\end{equation*}
and the Kepler's third law:
\begin{equation*}
    \omega^2 a^3 = g. 
\end{equation*}
Hence \eqref{L_dot} becomes:
\begin{equation}
    \label{L_dot_G_dot}
    \dot{L} = \frac{{{\dot E}^{(O)}}}{\omega} = \frac{\dot{\vartheta} + \dot{\varepsilon}}{\omega} \dot{G},
\end{equation}
where we have used \eqref{EO_dot}.\\
We now observe that several quantities, such as $\langle \dot\varphi\rangle=\Omega$, $\langle \dot\vartheta\rangle=\omega$ and both the orbital energy and the orbital angular momentum (and hence $L$ and $G$) varies very slowly. We can therefore assume that these quantities remain constant on one orbit and receive a very small increase (or decrease) at the end of each revolution.\\
Therefore we can compute the average on one orbit of $\dot{G}$ and $\dot{L}$, obtaining:
\begin{equation}
    \label{dotGmean}
    \langle \dot{G} \rangle = \langle \alpha \left( \dot{\varphi} - \dot{\vartheta} - \dot{\varepsilon} \right) \rangle = \alpha \left( \Omega - \omega \right)
\end{equation}
and
\begin{equation}
    \label{dotLmean1}
    \langle \dot{L} \rangle = \langle \frac{\dot{\vartheta} + \dot{\varepsilon}}{\omega} \dot{G} \rangle = \frac{\alpha}{\omega} \langle \left( \dot{\vartheta} + \dot{\varepsilon} \right) \left( \dot{\varphi} - \dot{\vartheta} - \dot{\varepsilon} \right) \rangle.
\end{equation}
But: 
\begin{equation*}
    \vartheta = \omega t + 2e \sin (\omega t) \Longrightarrow \dot{\vartheta} = \omega + 2e \omega \cos (\omega t).   
\end{equation*}
And (see \eqref{solution} for the definition of $A$, $B$ and $\delta$):
\begin{equation*}
    \varepsilon = A + eB \sin (\omega t + \delta) \Longrightarrow \dot{\varepsilon} = eB \omega \cos (\omega t + \delta).   
\end{equation*}
Therefore:
\begin{align*}
    \langle \dot{L} \rangle 
    & = \frac{\alpha}{\omega} \langle \left[ \omega + 2e \omega \cos (\omega t) + eB \omega \cos (\omega t + \delta) \right] \left[ \Omega - \omega - 2e \omega \cos (\omega t) - eB \omega \cos (\omega t + \delta)  \right] \rangle\\ 
    & = \frac{\alpha}{\omega} \left[ \omega \left( \Omega - \omega \right) + \omega e \left( \Omega - 2\omega \right) \langle 2 \cos (\omega t) + B \cos (\omega t + \delta) \rangle - \omega^2 e^2 \langle [ 2 \cos (\omega t) + B \cos (\omega t + \delta) ]^2 \rangle \right] \\ 
    & = \langle \dot{G} \rangle  + \alpha e \left( \Omega - 2\omega \right) \langle 2 \cos (\omega t) + B \cos (\omega t + \delta) \rangle - \alpha \omega e^2 \langle [ 2 \cos (\omega t) + B \cos (\omega t + \delta) ]^2 \rangle ,
\end{align*}
where we have used \eqref{dotGmean}.\\
The computation of the remaining averages leads to: 
\begin{equation*}
    \langle 2 \cos (\omega t) + B \cos (\omega t + \delta) \rangle = 0
\end{equation*}
and
\begin{align*}
    \langle [ 2 \cos (\omega t) - B \cos (\omega t + \delta) ]^2 \rangle 
    & = \langle 4 \cos^2 (\omega t) +4B\cos (\omega t)\cos (\omega t + \delta) + B^2 \cos^2 (\omega t + \delta) \rangle\\
    & = 2 +4B \langle \cos^2 (\omega t)\cos (\delta) - \sin (\omega t) \cos (\omega t) \sin (\delta) \rangle + \frac{B^2}{2}\\
    & = 2 +2B \cos(\delta)  + \frac{B^2}{2}.
\end{align*}\\
Hence:
\begin{equation}
    \label{dotLmean2}
    \langle \dot{L} \rangle = \langle \dot{G} \rangle - \alpha \omega e^2 \left( 2 +2B \cos(\delta)  + \frac{B^2}{2}  \right) = \langle \dot{G} \rangle -  e^2 C,
\end{equation}
with: $C=\alpha \omega \left( 2 +2B \cos(\delta)  + \frac{B^2}{2}  \right) > 0$.\\

Now, if we assume initially $L \sim G$, with $L > G$, we have:
\begin{equation}
\begin{aligned}
    \frac{d}{dt} \frac{G^2}{L^2} 
    & = \frac{2G\dot{G}L^2-2L\dot{L}G^2}{L^4}
    = \frac{2G\dot{G}L^2-2LG^2(\dot{G}-e^2C)}{L^4}
    = \frac{2G\dot{G}L^2-2G\dot{G}G^2 + 2LG^2e^2C)}{L^4}\\
    & = \frac{2G\dot{G}(L^2-G^2) + 2LG^2e^2C}{L^4} 
      = \frac{2G\dot{G}L^2e^2 + 2LG^2e^2C}{L^4} 
      = 2\frac{G\dot{G}L^2 + LG^2C}{L^4}e^2\\ 
    & = \frac{2}{\tau_M} e^2,
    \label{e_dot_2} 
\end{aligned}
\end{equation}
with $\tau_M = \frac{L^4}{G\dot{G}L^2 + LG^2C} > 0$.\\
Finally, the Delaunay variable $L$ ang $G$ satisfy the relation $G^2 = (1 - e^2) L^2$, hence we have:
\begin{equation}
    \label{e_dot_3} 
    \frac{d}{dt} \frac{G^2}{L^2} = \frac{d}{dt} (1-e^2) = -2e\dot{e}.  
\end{equation}
Putting together \eqref{e_dot_2} and \eqref{e_dot_3}, we obtain:
\begin{equation}
    \label{e_dot_4} 
    -2e\dot{e} = \frac{2}{\tau_M} e^2 \implies \dot{e} = - \frac{e}{\tau_M} \implies e(t) = e_0 \exp{\left(-\frac{t}{\tau_M}\right)}.  
\end{equation}
Therefore the eccentricity tends exponentially to zero for $t \to \infty$: the orbit becomes circular.

Consider now the limiting case of a circular orbit: $e=0$. In this case, on each single orbit $\dot{\varphi} = \Omega = const$ and $\dot{\vartheta} = \omega = const$ and then in equation \eqref{epsilon2} we have a constant forcing. Hence the solution of such equation is asymptotic to a constant: $\varepsilon = const$\\
From \eqref{epsilon2} and \eqref{G_dot} we then have:
\begin{equation}
    \label{G_dot2}
    \dot{G} = \alpha \left( \dot{\varphi} - \dot{\vartheta} \right) = \alpha \left( \Omega - \omega \right) = - \frac{\partial \mathcal{L}}{\partial \varepsilon} = \Gamma.
\end{equation}
We assume, as discussed before, that the very slow variation on $\Gamma$ is applied as a final kick after an unperturbed Keplerian orbit on which $\Gamma$ is assumed constant. We have:
\begin{equation}
    \label{G_dot_disc}
    G(T) - G(0) = \int_0^T \frac{dG}{dt} \,dt = \Gamma T.
\end{equation}
Hence, from \eqref{Gdef} and using ghe Kepler's third law $a^3\omega^2=g$, we have:
\begin{equation*}
    G(T) = G(0) + \Gamma T = m\sqrt{ga(0)} + \Gamma T,
\end{equation*}
and
\begin{equation*}
    G(T) = m\sqrt{ga(T)} = m \sqrt{g[a(0) + \dot{a} T]};
\end{equation*}
therefore:
\begin{equation*}
    m \sqrt{g[a(0) + \dot{a} T]} = m\sqrt{ga(0)} + \Gamma T \implies ga(0) + g\dot{a} T = ga(0) + \frac{2\Gamma}{m} T \sqrt{ga(0)} \implies \dot{a} = \frac{2\Gamma}{m} \sqrt{\frac{a(0)}{g}},
\end{equation*}
where we have neglected the term $\left( \frac{\Gamma}{m} T \right)^2$.\\
Finally, using again Kepler's third law:
\begin{equation}
    \label{a_dot}
    \dot{a} = \frac{2\alpha}{m\omega a} \left( \Omega - \omega \right).
\end{equation}
Note that in \eqref{a_dot} the dependence of $\dot a$ on $(\Omega-\omega)$ is regular (namely linear), while in literature, see \cite{MurrayDermott} (4.160), the dependence has a singularity for $(\Omega-\omega)=0$, since it depends on ${\rm sign}(\Omega-\omega)$. Note also that in this context the results one finds in literature corresponds to a different choice of the friction law in \eqref{frictional_force}, i.e. the choice $f=-const$. This choice is equivalent to considering the system as if it were made up of two solid surfaces that slide over each other.\\
We want to point out that we are not the first to notice this non-singularity, in fact there are several models in literature of tides in which a Rayleigh dissipation function is used and no singularity appears in the equation for $\dot{a}$, see for istance \cite{ragazzo2017viscoelastic} $(62)$. 

\section{JUPITER--IO SYSTEM.}

\begin{figure}[!ht]
    \includegraphics[width=0.7\textwidth]{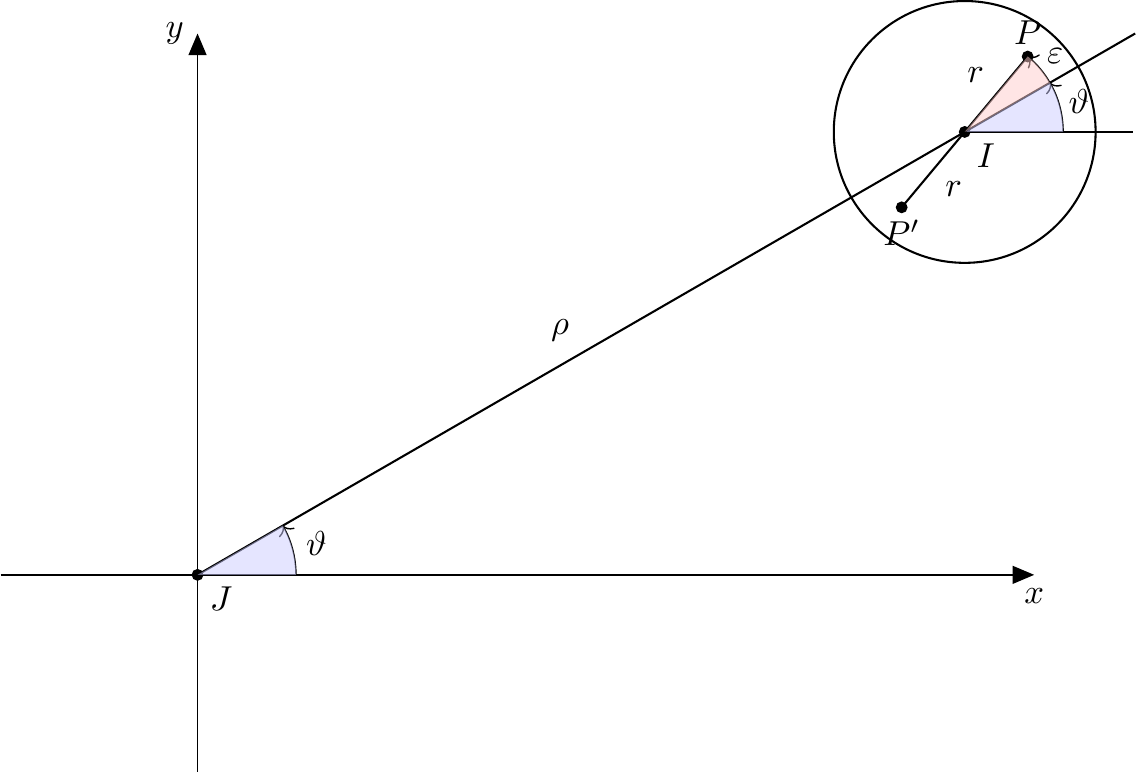}
    \caption{Jupiter--Io System}
    \label{fig:Jupiter_io_system}
\end{figure}

In this section we want to study the evolution of the Jupiter--Io system using the same formalism developed above.\\
To this end, as we show in Fig.~\ref{fig:Jupiter_io_system}, we imagine Io, of total mass $m$, as a sphere plus a symmetric dumbbell of diameter $2r$ and mass $\mu$. We imagine Jupiter, indicated with $J$, as a point of mass $M_J$ placed at the origin of the reference frame. We indicate with $\rho$ the distance between Jupiter and the center of Io: we observe that in an elliptical orbit this distance varies with time. Finally, we call $\varphi$, $\vartheta$ and $(\vartheta + \varepsilon)$ the angular positions of the rotation of Io, the angular positions of the revolution of Io and the angular position of the dumbbell with respect to a fixed direction ($x$-axis), respectively. So $\dot{\varphi}$, $\dot{\vartheta}$ and $(\dot{\vartheta} + \dot{\varepsilon})$ are the angular velocity of the rotation of Io, the angular velocity of the revolution of Io and the angular velocity of the dumbbell respectively.\\
In this case, the kinetic energy of this system is the sum of kinetic energies of the dumbbell and the kinetic energy of Io, that has two pieces: one due to the revolution around Jupiter and one due to the rotation around its own axis. The potential energy is the sum of three pieces of gravitational attraction: that between Io (deprived of the dumbbell) and Jupiter; that between point $P$ (a bulge of the dumbbell) and Jupiter and that between point $P'$ (the other bulge of dumbbell) and Jupiter. The potential can be expanded up to the second order in $\frac{r}{\rho}$ (that is a dimensionless small parameter).\\
So, the Lagrangian of the system is very similar to that of the previous section:
\begin{equation}
    \label{Lagrangian_Io}
    \mathcal{L} = \frac{1}{2} m (\dot{\rho}^2 + {\rho}^2 \dot{\vartheta}^2) + \frac{1}{2} I \dot{\varphi}^2 + \frac{1}{2} \mu r^2 (\dot{\vartheta} + \dot{\varepsilon})^2 + \frac{gm}{\rho} \left[1 + \frac{\mu}{m} \frac{r^2}{{\rho}^2} \left( \frac{3}{2} \cos^2{\varepsilon} - \frac{1}{2}\right) \right],
\end{equation}
where $I$ represents the moment of inertia of Io, while in \eqref{Lagrangian} it represents the moment of inertia of Earth. Moreover now, $g=kM_J$.\\ 
Assuming the same dissipation mechanism of previous section, justified by the fact that it is known that Io is made of molten material, we have the same Rayleigh's dissipation function of the previous section:
\begin{equation}
    \label{Rayleigh_Io}
    R=-\frac{1}{2}\alpha \left( \dot{\varphi} - \dot{\vartheta} - \dot{\varepsilon} \right)^2.
\end{equation}
So the equations of motion are the same of the previous section:
\begin{equation}
    \label{rho2_Io}
    m \Ddot{\rho} = \frac{\partial \mathcal{L}}{\partial \rho},
\end{equation}
\begin{equation}
    \label{theta2_Io}
    \frac{d}{dt} \left( m \rho^2 \dot{\vartheta} \right) = \alpha \left( \dot{\varphi} - \dot{\vartheta} - \dot{\varepsilon} \right),
\end{equation}
\begin{equation}
    \label{phi2_Io}
    I \Ddot{\varphi} = - \alpha \left( \dot{\varphi} - \dot{\vartheta} - \dot{\varepsilon} \right),
\end{equation}
\begin{equation}
    \label{epsilon2_Io}
    \frac{\partial \mathcal{L}}{\partial \varepsilon} + \alpha \left( \dot{\varphi} - \dot{\vartheta} - \dot{\varepsilon} \right) = 0.
\end{equation}\\
However in this case the initial conditions are different. In fact while in Earth--Moon system we have: $\dot{\varphi} \gg \dot{\vartheta}$ (in fact the period of rotation of Earth is much smaller than the period of revolution of Moon), in Jupiter--Io system we have: $ \langle \dot{\varphi} \rangle = \langle \dot{\vartheta} \rangle = \omega$ (in fact the period of rotation of Io is the same of his period of revolution around Jupiter).\\   
Equation \eqref{epsilon2_Io} admits a solution of the form: $\varepsilon (t) = e B \sin{(\omega t + \delta)}$, with $B$ and $\delta$ constants. This is evident if one performs all the steps seen in the appendix in the case of the Earth--Moon system.\\
Although $\varepsilon (t)$ is different from that of the previous section (in fact here it does not contain the constant term $A$), all the results up to \eqref{L_dot_G_dot} remain still valid.\\
The average on one orbit of $\dot{G}$ and $\dot{L}$ are slightly different from those in the previous section:
\begin{equation}
    \label{dotGmeanIo}
    \langle \dot{G} \rangle = 0
\end{equation}
and
\begin{equation}
    \label{dotLmean1Io}
    \langle \dot{L} \rangle = \langle \frac{\dot{\vartheta} + \dot{\varepsilon}}{\omega} \dot{G} \rangle = \frac{\alpha}{\omega} \langle \left( \dot{\vartheta} + \dot{\varepsilon} \right) \left( \dot{\varphi} - \dot{\vartheta} - \dot{\varepsilon} \right) \rangle.
\end{equation}
From now on, all the steps done in the previous section are the same, with $\langle \varphi \rangle = \omega$.
Therefore:
\begin{equation}
\begin{aligned}
    \langle \dot{L} \rangle
    & = \frac{\alpha}{\omega} \langle \left[ \omega + 2e \omega \cos (\omega t) + eB \omega \cos (\omega t + \delta) \right] \left[\omega - \omega - 2e \omega \cos (\omega t) - eB \omega \cos (\omega t + \delta)  \right] \rangle \\
    & = \frac{\alpha}{\omega} \left[ - \omega^2 e  \langle 2 \cos (\omega t) + B \cos (\omega t + \delta) \rangle - \omega^2 e^2 \langle [ 2 \cos (\omega t) + B \cos (\omega t + \delta) ]^2 \rangle \right]\\
    & = - \alpha \omega e^2 \langle 4 \cos^2 (\omega t) +4B\cos (\omega t)\cos (\omega t + \delta) + B^2 \cos^2 (\omega t + \delta) \rangle\\
    & = - \alpha \omega e^2 \left( 2 +2B \cos(\delta)  + \frac{B^2}{2}  \right) = -  e^2 C,
    \label{dotLmean2Io}
\end{aligned}
\end{equation}
with: $C=\alpha \omega \left( 2 +2B \cos(\delta)  + \frac{B^2}{2}  \right) > 0$.\\

Therefore:
\begin{equation}
    \label{e_dot_2Io}
    \frac{d}{dt} \frac{G^2}{L^2} = -\frac{2L\dot{L}G^2}{L^4}= \frac{2LG^2e^2C}{L^4} = \frac{2}{\tau_I} e^2, 
\end{equation}
with $\tau_I = \frac{L^4}{LG^2C} > 0$.\\
Finally:
\begin{equation}
    \label{e_dot_3Io} 
    \frac{d}{dt} \frac{G^2}{L^2} = \frac{d}{dt} (1-e^2) = -2e\dot{e}.  
\end{equation}
Putting together \eqref{e_dot_2Io} and \eqref{e_dot_3Io}, we obtain:
\begin{equation}
    \label{e_dot_4Io} 
    -2e\dot{e} = \frac{2}{\tau_I} e^2 \implies \dot{e} = - \frac{e}{\tau_I} \implies e(t) = e_0 \exp{\left(-\frac{t}{\tau_I}\right)}.  
\end{equation}
Therefore the eccentricity tends exponentially to zero for $t \to \infty$: the orbit becomes circular even in Jupiter--Io case.

\section{CONCLUSION}
In this short note we proposed a simplified model to describe the effects of tidal dissipation on the orbital parameters of the two body problem. We think that the main virtue of the model lies in the clarification of the relevance of the model of friction used in order to describe the interaction between the body and the tidal bulges. We have pointed out that some of the classical results assume tacitly that this friction is velocity independent, like the solid-on-solid friction, while a viscous friction seems to be more realistic and solves some difficulties present in the classical theory.

\section*{APPENDIX A. COMPUTATION OF $\varepsilon (t)$}
\setcounter{equation}{0}
\renewcommand{\theequation}{{\rm A}.\arabic{equation}}

In this appendix we want to show the detailed computation of $\varepsilon (t)$ starting from equation \eqref{epsilon2} that we rewrite in this form:
\begin{equation}
    \label{epsilon_eq}
    \alpha \dot{\varepsilon}(t) - \frac{\partial \mathcal{L}}{\partial \varepsilon} - \alpha \left( \dot{\varphi} - \dot{\vartheta} \right) = 0.
\end{equation}\\
First of all, we compute $\frac{\partial L}{\partial \varepsilon}$ from equation~\eqref{Lagrangian}:
\begin{equation}
    \label{Gamma}
    \frac{\partial \mathcal{L}}{\partial \varepsilon} =- \frac{gm}{\rho} \frac{\mu}{M_E} \frac{r^2}{{\rho}^2} 3 \cos{\varepsilon} \sin{\varepsilon} \simeq -3 \frac{k m\mu}{\rho} \frac{r^2}{{\rho}^2} \varepsilon,
\end{equation}\\
where we have used $ g = k M_E $ after developing $\cos{\varepsilon} \sin{\varepsilon}$ in power series of $\varepsilon$ and keeping only the linear terms in $\varepsilon$.\\
Recall that
$\mu$ represents the mass of the ocean's bulges. 
These bulges can be imagined 
as an ellipsoid of radii $R_E$, $R_E$ and $R_E+h$ deprived of a sphere of radius $R_E$ concentric to it, being $h$ the tidal height.\\
If we replace $h$ with the Newton formula for tidal height:
\begin{equation}
    h=\frac{3}{2}\frac{m}{M_E} \left( \frac{R_E}{\rho} \right)^3 R_E,
\end{equation}
then $\mu$ becomes:
\begin{equation}
    \label{mu}
    \mu = \delta_W \frac{4}{3} \pi R_E^2 h = \delta_W \frac{4}{3} \pi R_E^2 \frac{3}{2}\frac{m}{M_E} \left( \frac{R_E}{\rho} \right)^3 R_E = \frac{3}{2}\frac{\delta_W}{\delta_E} m \left( \frac{R_E}{\rho} \right)^3w
\end{equation}\\
where $\delta_W$ and $\delta_E$ represent the densities of the liquid part and the solid part of the planet.
Finally, in \eqref{Gamma}, $r$ is the distance between the center of the Earth and the center of mass of 
each
bulge, that is the center of mass of the aforementioned ellipsoid deprived of a sphere. It is a standard calculation to show that,
$r=\frac{3}{4}R_E$ up to terms that go to zero as $\frac{h}{R_E}$.\\
Therefore:
\begin{equation}
    \label{Gamma_2}
    \frac{\partial \mathcal{L}}{\partial \varepsilon} = -\frac{81}{32} \frac{\delta_W}{\delta_E} k m^2 \frac{R_E^5}{{\rho}^6} \varepsilon = -D k m^2 \frac{R_E^5}{{\rho}^6} \varepsilon,
\end{equation}
with $D = \frac{81}{32} \frac{\delta_W}{\delta_E}$ a dimensionless constant.\\
Equation \eqref{epsilon_eq} 
thus
becomes:
\begin{equation}
    \label{epsilon_eq_2}
    \alpha \dot{\varepsilon} + D k m^2 \frac{R_E^5}{{\rho}^6} \varepsilon - \alpha \left( \dot{\varphi} - \dot{\vartheta} \right) = 0.
\end{equation}\\
Finally, we have to consider the time-dependence of $\rho$ and $\vartheta$. We also assume that $\Omega$, $\omega$ and $a$ remain constant during each revolution and 
change their values
only at the end of each revolution. 
This assumption yiedls
\begin{equation}
    \label{theta_time}
    \vartheta \simeq \lambda + 2e\sin{\lambda} =  \omega t + 2e\sin{(\omega t)} \implies \dot{\vartheta} \simeq \omega + 2\omega e \cos{(\omega t)} 
\end{equation}
and 
\begin{equation}
    \label{rho_time}
    \rho (t) = \frac{p}{1+e \cos{(\omega t)}} = \frac{a(1-e^2)}{1+e \cos{(\omega t)}} \simeq a [1-e \cos{(\omega t)}] \implies \frac{1}{\rho^6} \simeq \frac{1}{a^6} [1+6e \cos{(\omega t)}]. 
\end{equation}\\
So, equation \eqref{epsilon_eq_2} becomes:
\begin{equation}
    \label{epsilon_eq_3}
    \dot{\varepsilon} + \frac{\gamma_c}{\alpha} [1+6e \cos{(\omega t)}] \varepsilon - \Omega + \omega + 2 \omega e \cos(\omega t)= 0,
\end{equation}
where $\gamma_c =  D k m^2 \frac{R_E^5}{{a}^6}$.\\
A trivial solution of this equation is:
\begin{equation}
    \label{solution}
    \varepsilon(t) = \frac{\alpha}{\gamma_c} (\Omega - \omega) - e \frac{6\Omega -4\omega}{\omega} \cos{\delta} \sin (\omega t + \delta) = A + eB \sin (\omega t + \delta), 
\end{equation}
with $A = \frac{\alpha}{\gamma_c} (\Omega - \omega), B = \frac{6\Omega -4\omega}{\omega} \cos{\delta} \text{ and} \tan \delta = \frac{\gamma_c}{\alpha \omega}$.

{\it Remark}: the ratio $\frac{\alpha}{\gamma_c}$ contains, as a whole, the information about the quantities appearing in the system. This suggests for instance that the friction coefficient $\alpha$ has to decay very fast for increasing $a$, the semi-major axis of the orbit, in order to balance the dependance on $a$ of $\gamma_c$ (see the comments immediately following~\eqref{frictional_force}). This sounds reasonable, since the friction should depend on the total amount of liquid involved in the motion of the bulges, and this clearly depends on $a$. However we do not have a detailed model of the tidal currents inside the oceans (in the Earth-Moon case) or in the inner mantle (Io-Jupiter case). We think that one of the virtues of the simple model we presented in this work is the fact that it suggests the correct relations among the various elements of the orbits, while their quantitative evaluation has to be based on empirical observations.

\section*{APPENDIX B. CHARACTERISTIC TIMESCALES OF THE SYSTEM}
\label{sec:characteristic_timescalse}
\setcounter{equation}{0}
\renewcommand{\theequation}{{\rm B}.\arabic{equation}}

In this appendix we want to compare the characteristic timescales of the system in order to show that the assumption we made in our model are justified.\\
In particular, as we explained above, we are interested in the comparison between the timescale of elongation of semi-major axis, that we call $\tau_a$ and the timescale of circularization of the orbit, that is $\tau_M$ or $\tau_I$ in equation \eqref{e_dot_4} and \eqref{e_dot_4Io} respectively.\\
From equation \eqref{a_dot} we see that $\tau_a \simeq \tau_{\Omega} $, with $\tau_{\Omega}$ the characteristic timescale of evolution of $\Omega$. Moreover, we can rewrite equation \eqref{phi2} as
\begin{equation}
    I\dot{\Omega} \simeq -\alpha \Omega
\end{equation}
hence we have $\tau_{\Omega} \simeq \frac{I}{\alpha}$.\\
On the other hand:
\begin{equation}\label{eq:tauM}
    \tau_M = \frac{L^4}{G\dot{G}L^2 + LG^2C} \simeq \frac{L}{\dot{G} + C} \simeq \frac{L}{\alpha}\frac{1}{18\Omega \frac{\Omega}{\omega}}.
\end{equation}
Therefore, in the case of the Earth-Moon system, and, hence
$\tau_M \simeq \frac{L}{500 \alpha \Omega}$
\begin{equation}
     \frac{\tau_a}{\tau_M} \simeq
     \frac{500 I \Omega}{L} \simeq
     500\frac{\Omega}{\omega} \frac{\frac{2}{5}M_E R^2_E}{m R^2_{EM}} 
     \simeq 500 \cdot 28 \cdot \frac{2}{5} \frac{M_E}{m}  
     \left( \frac{R_E}{R_{EM}} \right)^2 \simeq 120 > 1.
\end{equation}
Conversely, in the case of the Jupiter-Io system,
the analogue of equation~\eqref{eq:tauM} is
$\tau_I \simeq \frac{L}{80 \alpha \Omega}$
yielding
\begin{equation}
     \frac{\tau_a}{\tau_I} \simeq
     \frac{80 I \Omega}{L} \simeq
     80 \frac{\Omega}{\omega} \frac{\frac{2}{5}M_J R^2_J}{m R^2_{JI}} = 32 \cdot 1 \cdot \frac{2}{5} \frac{M_J}{m} \left( \frac{R_J}{R_{JI}} \right)^2 
     \simeq 1.9 \times 10^4 \gg 1.
\end{equation}
These calculations show that the assumption of a constant $\mu$
appears to be very reasonable  already
in the case of the Earth-Moon system.
In the case of Jupiter and Io the approximation is
much better and the difference between a variable and a constant $\mu$ is
really negligible.

\section*{ACKNOWLEDGMENTS}
The authors want to thank the anonymous referees whose
contribution helped to improve the readability of the paper.
We are indebted to Ugo Locatelli for many useful discussions. We benefited of several comments by Giuseppe Pucacco and Gabriella Pinzari.
This work has been supported by PRIN-CELMECH. 
BS acknowledges the support of the Italian MIUR Department of Excellence grant (CUP E83C18000100006).
AT has been supported through the H2020 Project Stable and Chaotic Motions in the Planetary Problem (Grant 677793 StableChaoticPlanetM of the European Research Council)

\printbibliography
\end{document}